\documentclass[review]{elsarticle}

\usepackage{lineno}

\usepackage{multirow,setspace,amssymb,amsmath,graphicx,color,rotating,subfigure,url}
\usepackage{lineno}
\usepackage{textcomp}
\usepackage{booktabs}    %导言区	
\usepackage{CJK}
\usepackage{bm}%
\usepackage{rotating}
\usepackage{diagbox}
\usepackage{epsfig}% Include figure files
\usepackage{dcolumn}% Align table columns on decimal point
\usepackage{epstopdf}
\usepackage{natbib}
\biboptions{sort&compress}
\usepackage[hidelinks]{hyperref}
\usepackage[title]{appendix}%

\begin{document}

\begin{frontmatter}

\title{A hyper-distance-based method for hypernetwork comparison}

\author[inst1]{Tao Xu}
\author[inst1]{Xiaowen Xie}
\author[inst2,inst3]{Zi-Ke Zhang}
\author[inst1]{Chuang Liu}
\author[inst1,inst3]{Xiu-Xiu Zhan\corref{ca1}}\ead{zhanxiuxiu@hznu.edu.cn}
%\ead{liuchuang@hznu.edu.cn}
\address[inst1]{Alibaba Research Center for Complexity Sciences, Hangzhou Normal University, Hangzhou, 311121, P. R. China}
\address[inst2]{Digital Communication Research Center, Zhejiang University, Hangzhou 310058, China}
\address[inst3]{College of Media and International Culture, Zhejiang University, Hangzhou 310058, PR China}

\cortext[ca1]{Corresponding authors.}

\begin{abstract}
% Identifying vital nodes in the complex network has been a crucial topic in the field of network science.
Hypernetwork is a useful way to depict multiple connections between nodes, making it an ideal tool for representing complex relationships in network science.
In recent years, there has been a marked increase in studies on hypernetworks, however, the comparison of the difference between two hypernetworks has been given less attention.
This paper proposes a hyper-distance-based method (HD) for comparing hypernetworks. This method takes into account high-order information, such as the high-order distance between nodes. The experiments carried out on synthetic hypernetworks have shown that HD is capable of distinguishing between hypernetworks generated with different parameters, and it is successful in the classification of hypernetworks. Furthermore, HD outperforms current state-of-the-art baselines to distinguish empirical hypernetworks when hyperedges are disrupted.

\end{abstract}


\end{frontmatter}

%\linenumbers
\section{Introduction}
\label{Intro}
\makeatletter
\newcommand{\rmnum}[1]{\romannumeral #1}
\newcommand{\Rmnum}[1]{\expandafter\@slowromancap\romannumeral #1@}
\makeatother

With the rapid development of the Internet and the explosive growth of information ~\cite{boccaletti2006complex}, the size and complexity of the network data continue to increase. In the modern digital era, it is essential to compare and evaluate the distinctions between various networks~\cite{tantardini2019comparing} in order to gain a thorough understanding of the structure, function, and behavior of the network. The exploration of network comparison is a significant area of study with implications for disciplines such as network science, data mining, and machine learning~\cite{albert2002statistical,hand2007principles,rolnick2022tackling,xu2023reconstruction} and can be used to solve problems such as network classification, algorithm evaluation, and anomaly detection~\cite{fkih2022similarity, tabak2019machine,greener2022guide}. 
Nevertheless, the customary techniques for contrasting networks mainly concentrate on the low-order features of networks. For instance, 
Han et al.~\cite{hand2007principles} proposed a straightforward and effective measure, i.e., the Jaccard method, which calculates the difference between the adjacency matrices of two networks. Koutra et al.~\cite{koutra2016deltacon} employed a method called Deltacon to compare the similarity between all pairs of nodes in two networks. Aliakbary et al.~\cite{aliakbary2015distance} analyzed the graphlet distribution of a network and compared the similarity between two networks through the graphlet distribution.
 Furthermore, Bagrow et al.~\cite{bagrow2019information} introduced the network portrait and KL divergence to compare two networks.
 Despite the fact that these techniques may partially uncover the fundamental characteristics of networks~\cite{wang2022quantification}, they have restrictions on grasping the intricate multivariate connections existing in complex systems.
% {\color{red}stop}
In the real world, many complex systems exhibit rich multivariate relationships, including dependencies among multiple variables and interactions among multiple objects. For example, in collaboration networks, multiple individuals may collaborate on one project.
In the brain's neural system, multiple regions can be highly active simultaneously. In social platforms, groups made up of multiple individuals share information and interact with each other~\cite{newman2002structure,battiston2020networks,benson2016higher,sweeney2021network}. The relationships mentioned above can be modeled by hypernetworks, which are networks composed of nodes and hyperedges. A hyperedge may contain more than two nodes, and a single node may belong to multiple hyperedges ~\cite{xie2023efficient}. Representing a complex system as a hypernetwork can break the limitation of "binary interactions" in an ordinary network and better reflect the multidimensional interactions present in the system ~\cite{chodrow2021generative}. Several methods have been proposed to compare differences in hypernetworks, such as those of Surana et al.~\cite{surana2022hypergraph}. However, most of these methods are adaptations of ordinary network comparison methods and do not fully utilize the high-order characteristics specific to hypernetworks.

To address this problem, we propose a hyper-distance-based network comparison method (HD), which reveals a comprehensive picture of network dissimilarity and characterizes the dissimilarity between hypernetworks using high-order information. The method compares the differences between two hypernetworks through the shortest path lengths defined by the high-order distance. 
We apply the method to compare hypernetworks generated by network models as well as empirical hypernetworks deduced by real data from different domains. The results show that our method can successfully compare different networks and perform better than the baselines.

The rest of this paper is structured as follows. Section 2 outlines the specifics of the hyper-distance-based network comparison approach. Section 3 provides an overview of the baseline methods. Section 4 presents the experimental results of comparing different hypernetworks. Finally, Section 5 summarizes the findings and suggests potential future research directions.

\section{Hyper-distance-based network comparison method}
\label{sec:Hypergraph and Datasets}
In this section, we propose a hypernetwork comparison method based on the local and global properties of hypernetworks. We first illustrate the basic definitions including the high-order distance, and the hypernetwork comparison method is further defined based on the high-order distance.
\subsection{Basic Definitions}
\label{sec:Definition of a Hypergraph}
A hypernetwork is defined as $H = (V, E)$, with sets of nodes $V$ and hyperedges $E$ represented as $V = \{v_1, v_2, \cdots, v_N\}$ and $E = \{e_1, e_2, \cdots, e_M\}$, respectively, where $N$ and $M$ represent the number of nodes and hyperedges. An incident matrix $I_{N*M}$ is constructed based on the membership relationship between the nodes and hyperedges. If node $v_i$ is part of a hyperedge $e_j$, then the value of $I_{ij}$ is set to $1$, otherwise it is $0$. Therefore, we can obtain the following.

\begin{equation}
    k^{H}(v_i)=\sum_{e_j\in{E}}I_{ij},
\end{equation}

\begin{equation}
    k^{E}(e_j)=\sum_{v_i\in{V}}I_{ij},
\end{equation}
where $k(v_i)$ and $k^{E}(e_j)$ represent the hyperdegree of node $v_i$ and the size of hyperedge $e_j$, respectively.

We create the ordinary network $G=(V, E^{o})$ of $H$ by linking all the nodes that are part of a hyperedge in $H$. In other words, an edge $e^{o}=(v_i,v_j) \in E^{o}$ if $v_i$ and $v_j$ both appear in at least one of the hyperedges in $E$. The network $G$ is an unweighted and undirected network, and its adjacency matrix is denoted as $A^{o}$. This means that if there is an edge between the nodes $v_i$ and $v_j$ in $G$, then $A^{o}_{ij}$ is equal to 1, otherwise it is 0.

\textbf{The $s$-distance between nodes: } We employ the hyperedge distance from~\cite{aksoy2020hypernetwork} to calculate the $s$-distance between two nodes. Two hyperedges are considered to be $s$-adjacent if they have at least $s$ nodes in common. Concretely, the $s$-distance between two hyperedges, $e_i$ and $e_j$, is denoted as $L_s^E(e_i,e_j)$ and is expressed as:

\begin{equation}\small
\label{NodeDistance}
    L_s^E(e_i,e_j)=
                    \begin{cases} 
                    \text{The shortest }s\text{-path length,} & \text {if there exists an }s\text{-path between }e_i\text{ and }e_j \\
                    \infty, & \text {if there is no }s\text{-path between}e_i\text{ and }e_j
                    \end{cases}.
\end{equation}

Based on the above definition, the $s$-distance between two nodes $v_i$ and $v_j$ is $L_s^V(v_i, v_j)$ and can be defined as follows:

\begin{equation}\small
\label{NodeDistance}
    L_{s}^{V}(v_i, v_j)=
                    \begin{cases} 
                    1, & \text {if } v_i,\ v_j \text{ exist in the same hyperedge} \\
                    \min(L_s^E(e_m,e_n)) + 1,  & v_i\in{e_m},v_j\in{e_n}
                    \end{cases}.
\end{equation}

\subsection{Hyper distance-based method for hypernetwork comparison} 
\textbf{The $s$-distance distribution of a hypernetwork: }
The proportion of nodes whose $s$-distance from node $v_i$ is denoted by $p_i^s(j)$. The $s$-distance distribution of node $v_i$ is represented by a vector of size $1\times(d^s+1)$, denoted as $P_i^s=\{p_i^s(j)\}$, where $d^s$ is the maximum value of $s$-distance.
 The proportion of nodes that have an $s$-distance of infinity from node $v_i$ is represented by $p_i^s(d^s+1)$. Consequently, we define the $s$-distance distribution of the hypernetwork $H$ as $P^s=\{P_1^s,P_2^s,\cdots,P_N^s\}$, indicating the detailed topological information with respect to the $s$-distance between nodes. When $s=1$, the distance distribution of the ordinary network is obtained, which is derived by $H$, is obtained; when $s>1$, a high-order distance distribution is revealed, which reflects the various high-order topological relationships between nodes that are indicated by different values of $s$. It should be noted that for a hypernetwork $H$, we have $1\leq s \leq S_m$, where $S_m$ represents the maximum number of overlapping nodes between hyperedges.

\textbf{Dissimilarity between two hypernetworks}: Given a hypernetwork $H$ and its $s$-distance distribution $P^s=\{P_1^s, P_2^s,\cdots, P_N^s\} (1\leq s \leq S_m)$, we leverage the Jensen-Shannon divergence to define the $s$-order heterogeneity, which is named $s$-order hypernetwork node dispersion $NND^s(H)$.
The definition of $NND^s$ can be expressed as follows:

\begin{equation}\label{equ:PixelColorContrast}
    NND^s(H)=\frac{J(P_1^s,P_2^s,\cdots,P_N^s)}{\ln(d^s+1)},
\end{equation}
where $J(P_1^s,P_2^s,\cdots,P_N^s)$ represents the Jensen-Shannon divergence of nodes $s$-distance distribution, and is defined as:
\begin{equation}\label{equ:PixelColorContrast}
    J(P_1^s,P_2^s,\cdots,P_N^s)=\frac{1}{N}\sum_{i,j}p_i^s(j)\ln(\frac{p_i^s(j)}{{\mu}_j^s}),
\end{equation}
where ${\mu}_j^s$ denotes the average of the $N$ $s$-distance distributions, which is given by:

\begin{equation}\label{equ:PixelColorContrast}
   {\mu}_j^s=(\sum_{i=1}^Np_i^s(j))/N.
\end{equation}

We compare the dissimilarity between two hypernetworks, i.e., $H$ and $H^{'}$, based on the $s$-order heterogeneity defined above. Specifically, the structural dissimilarity $D(H,H^{'})$ is written as follows:

\begin{equation}\label{dissimilarityequation}\scriptsize
   D(H,H^{'})=\beta\sum_{s=1}^{S_m}\sqrt{\frac{J(\mu_H^s,\mu_{H'}^s)}{\ln2}}
   +(1-\beta)\sum_{s=1}^{S_m} \vert \sqrt{NND^s(H)}-\sqrt{NND^s(H^{'})} \vert ,
\end{equation}
where $J(\mu_H^s,\mu_{H'}^s)$ denotes the Jensen-Shannon divergence of the average $s$-distance distributions between $H$ and $H^{'}$.
 Dissimilarity $D(H, H^{'})$ is composed of two terms: The first term compares the average connectivity of nodes at different $s$-distances, capturing global dissimilarity between two hypernetworks; The second term integrates the dissimilarities among different network node dispersions via varying the value of $s$, indicating the local differences between two hypernetworks. We use a parameter $\beta (0<\beta<1)$ to adjust the importance of these two terms. A higher value of $D(H, H^{'})$ indicates larger differences between $H$ and $H^{'}$.

\begin{figure}[htp]
    \centering
    \includegraphics[width=1.1\columnwidth]{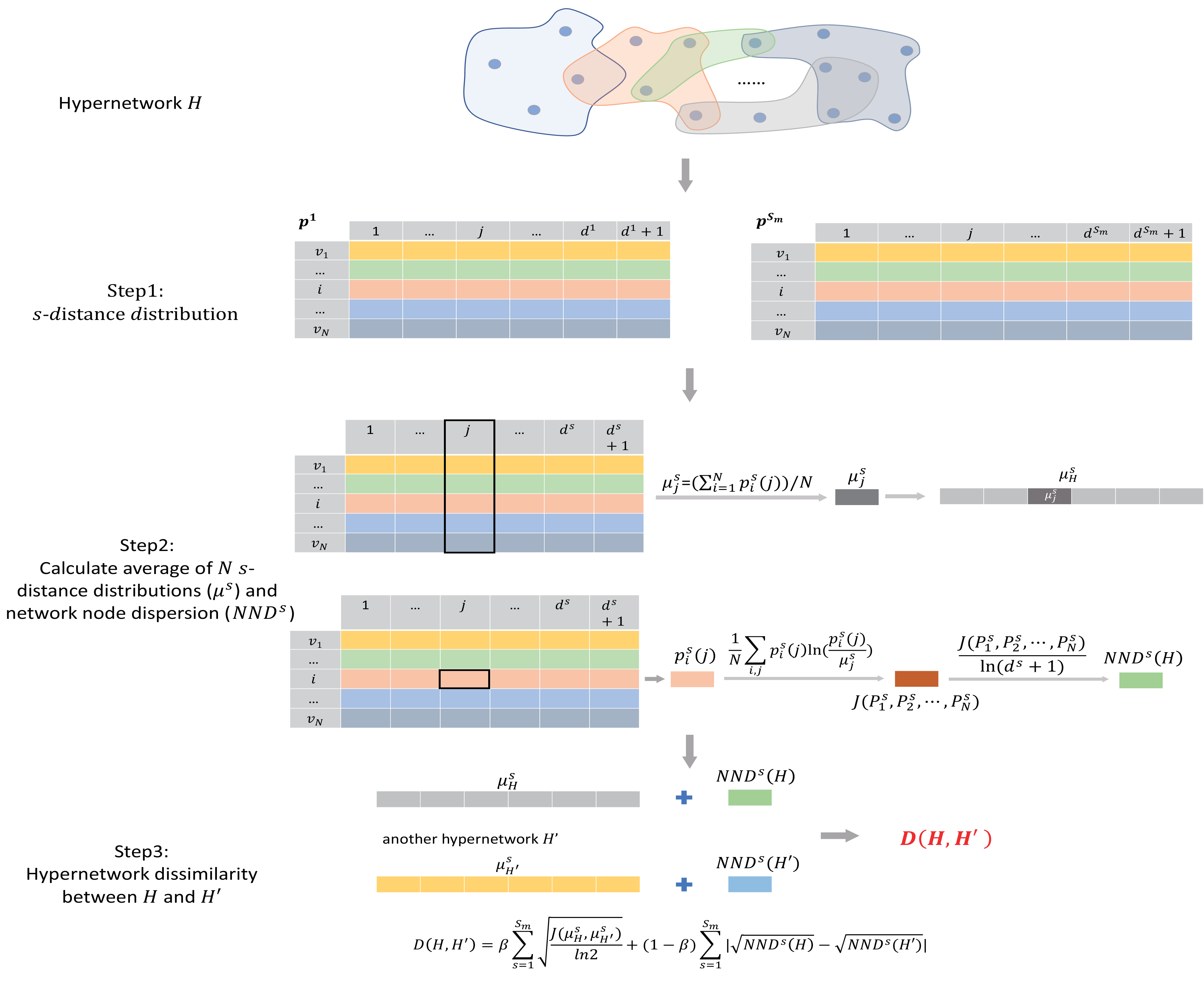}
    \caption{Flow diagram of hyper-distance-based method for hypernetwork comparison.}
    \label{fig:Di}
\end{figure}

 For clarity, we give a flow diagram of the proposed hypernetwork comparison method in Figure~\ref{fig:Di}. Taking a hypernetwork $H$ as an example, we first compute the $s$-distance distribution ($P^1, P^2, \cdots, {P^{S_m}}$). Subsequently, the values of $NND^s(H)$ and $\mu_H^s$ are calculated using Eqs.~(5-7). Consequently, the values of $NND^s(H^{'})$ and $\mu_{H^{'}}^s$ can also be computed. We further obtain the dissimilarity between $H$ and $H^{'}$ following Eq.~(\ref{dissimilarityequation}).

\clearpage
\section{Baseline Methods}
\label{Model Description}
Previous research~\cite{surana2022hypergraph} has largely relied on the projected ordinary network when comparing hypernetworks, meaning that two hypernetworks are first converted into ordinary networks and then the methods used for ordinary networks are applied for comparison.
We assess the efficacy of our approach by extending the comparison methods of ordinary networks to hypernetworks, including an information-theoretic approach based on portraits, adjacency tensors, and hypernetwork centrality-based methods. The specifics of each method are outlined below.
\subsection{Portrait-based hypernetwork comparison method}
For a given value of $s$, the s-distance-based hypernetwork portrait $B^{s}$ is an array with $(l,k)$ elements~\cite{bagrow2019information,bagrow2008portraits}, and we have
\begin{equation}
   B_{l,k}^s\equiv\mbox{the number of nodes who have }k\mbox{ nodes at s-distance }l,
\end{equation}
where $0\leq{l}\leq{d^s}$, $0\leq{k}\leq{N-1}$, and $d^s$ denotes the maximum $s$-distance value. It is noteworthy that the hypernetwork portrait $B^s$ is independent of the ordering or labeling of nodes. We use probability distribution $Q^s$ to comprehend the entries of $B^{S}$.  Thus, the probability of a randomly selected node having $k$ nodes at a distance of $l$ is expressed by
\begin{equation}
   Q^s(l \vert k)=\frac{1}{N}B^s_{l,k}.
\end{equation}

We consider two hypernetworks, $H$ and $H^{'}$, and use $Q^{s}$ and $Q^{s'}$ as probability distributions to interpret the rows of the $s$-distance-based hypernetwork portraits for each of them. The difference between $H$ and $H^{'}$ is expressed by $D^P(H, H')$ and is stated as:
\begin{equation}
    D^p(H,H^{'})=\sum_{s=1}^{S_m}(\frac{1}{2}KL(Q^s \Vert M^s)+\frac{1}{2}KL(Q^{s'} \Vert M^s)),
\end{equation}
where $M^s=\frac{1}{2}(Q^s+Q^{s'})$, $KL(\cdotp \vert \cdotp)$ is the Kullback-Liebler divergence~\cite{van2014renyi} between the two distributions. We compare the difference between $H$ and $H^{'}$ by comparing the difference of the portrait distributions of every order of distance with the alteration of $s$.

\subsection{Tensor-based hypernetwork Comparison Method}

The adjacency tensor of a hypernetwork $H$ is indicated by $T \in \mathbb{R}^{\xi \times {N} \times {N}}$~\cite{banerjee2017spectra,li2013z,chang2020hypergraph}, where $\xi$ is the maximum size of the hyperedges and $N$ is the number of nodes. If nodes $v_i$ and $v_j$ are both present in a hyperedge of size $k$, then $T_{kij}$ is set to 1; if not, it is 0. Subsequently, we use the adjacency tensor $T$ to compare the similarity between two hypernetworks $H$ and $H'$, which is represented as follows:

\begin{equation}
   D_T(H,H^{'})=\sum_{k=2}^{\xi}\sum_{i,j=1}^{N}\frac{1}{\gamma^k}|T_{kij}-T^{'}_{kij}|,
\end{equation}
where $\gamma$ is a parameter that can be tuned, and we choose $\gamma=2$ in the subsequent experiments.

\subsection{Centrality-based hypernetwork Comparison Methods}

To perform hypernetwork comparisons, we construct centrality vectors for nodes or hyperedges based on various centrality measures. The centrality of each node or hyperedge is denoted by $c = (c_1, c_2, \dots, c_N)^T$ or $c = (c_1, c_2, \dots, c_M)^T$ respectively, with $N$ and $M$ being the total number of nodes and hyperedges. The formula to compare two hypernetworks $H,H^{'}$ using node centrality methods can be written as:

\begin{equation}
   D_c(H,H')=\sum_{i=1}^N|c_i-c^{'}_i|,
\end{equation}
 and the formula for comparing $H,H^{'}$ using hyperedge centrality methods is given  by

 \begin{equation}
   D_c(H,H')=\sum_{i=1}^M|c_i-c^{'}_i|
\end{equation}
We utilize six centrality measures ~\cite{tantardini2019comparing,kovalenko2022vector,aksoy2020hypernetwork,hu2021aging,wang2010evolving,xie2023vital}, including node centrality methods such as Hyper Degree Centrality (HDC), Normalized Degree Centrality (NDC), Vector Centrality (VC) and three hyperedge centrality methods, s-Eccentricity Centrality (ECC), Edge-base Degree Centrality (EDC), s-Harmonic Closeness Centrality (HCC), to compare hypernetworks. Centrality measures are employed to create centrality vectors for nodes or hyperedges, which are then used as characteristics for the comparison of hypernetworks.

\section{Results}
\label{Results}
\subsection{Comparison of synthetic hypernetworks}
% 该部分将使用本文提出的方法和一些基准方法对不同参数下的合成网络进行对比，以此评估区分不同结构特征超网络的有效性。具体而言，即在同一拓扑结构超网络中，在不同的参数设置下，良好的超网络差异性比较方法会随着参数的变化而体现出显著的差异性（具体的合成网络构建方法见补充材料）。

We evaluate the performance of our approach by contrasting the synthetic uniform hypernetworks created by the Erdos-Rényi hypernetwork model~\cite{surana2022hypergraph}, Watts-Strogats hypernetwork model (WSH)~\cite{watts1998collective}, and the Scale-Free hypernetwork model (SFH)~\cite{ko2022growth,barabasi1999emergence}.
Details of how to generate synthetic hypernetworks are given below:

\textbf{Erdos-Rényi hypernetwork model (ERH).} The ERH model constructs a $k^{E}$-uniform hypernetwork $H=(V,E)$ given the number of nodes $N$ and the number of hyperedges $M$.
Specifically, we first randomly select $k^{E}$ nodes to form a hyperedge $e$ and add $e$ to $E$ if $e\notin E$, otherwise we reselect the nodes. We repeat the above process for $M$ times to obtain $H$.

%（2）Scale Free 超网络($SFH$)\cite{barabasi1999emergence}：给定节点数$N$,超边数$M$和斜率$r$来构造$k$-均匀$SF$超网络：
\textbf{Watts-Strogats hypernetwork model (WSH).}  The WSH model generates the $k^{E}$-uniform WS hypernetwork given the number of nodes $N$ and rewiring probability $q$. The model is described as follows:
Initially, a $k^{E}$-ring hypernetwork $H=\{N, E\}$ is given, where all nodes are arranged in a circular manner, and each node forms two hyperedges with its $k^{E}-1$ left and right nodes separately. For each hyperedge $e$, we randomly select $k^{E}$ nodes to form a new hyperedge $e^{'}$. If $e^{'} \notin E$, we replace $e$ by $e^{'}$ with probability $q$. The process will be terminated until all the hyperedges in $H$ are traversed.

% \textbf{Watts-Strogats hypernetwork model (WSH).}  The WSH model generates a $k$-uniform hypernetwork $H=(V,E)$ given the number of nodes $N$ and rewiring probability $p$. The model is described as follows: (i) Given a cyclic hypernetwork containing $N$ nodes, each node $v_i$ selects $k-1$ of its $w$ neighboring nodes randomly to form a hyperedge until there is at least one hyperedge containing $v_i$ and each of its neighbor separately, where $w$ is an even number; (ii) Select an initial hyperedge and generate a new hyperedge (randomly select $k$ nodes) If the same hyperedge does not exist in the original network then replace the selected initial hyperedge with the new hyperedge with $p$ probability; (iii) Repeat step ii until all hyperedge $e \in E$ have been traversed once.

\textbf{Scale-free hypernetwork model (SFH)}. Given the number of nodes $N$, the number of hyperedges $M$ and the degree distribution $p(k)\sim (k)^{-r}$, we generate a $k^{E}$-uniform SFH hypernetwork through the following steps: (i) we create a node degree sequence based on the degree distribution $p(k)\sim (k)^{-r}$, and each node $v_i$ is assigned a probability $p_i$ according to the node degree sequence; (ii) For an empty hyperedge $e$, we add node $v_i$ to $e$ with probability $p_i$ until the hyperedge size of $e$ reaches $k^{E}$. If $e$ already exists in $H$, we regenerate $e$; (iii) Repeat step (ii) for $M$ times.

\begin{figure}[!htp]
    \centering
    \includegraphics[width=\columnwidth]{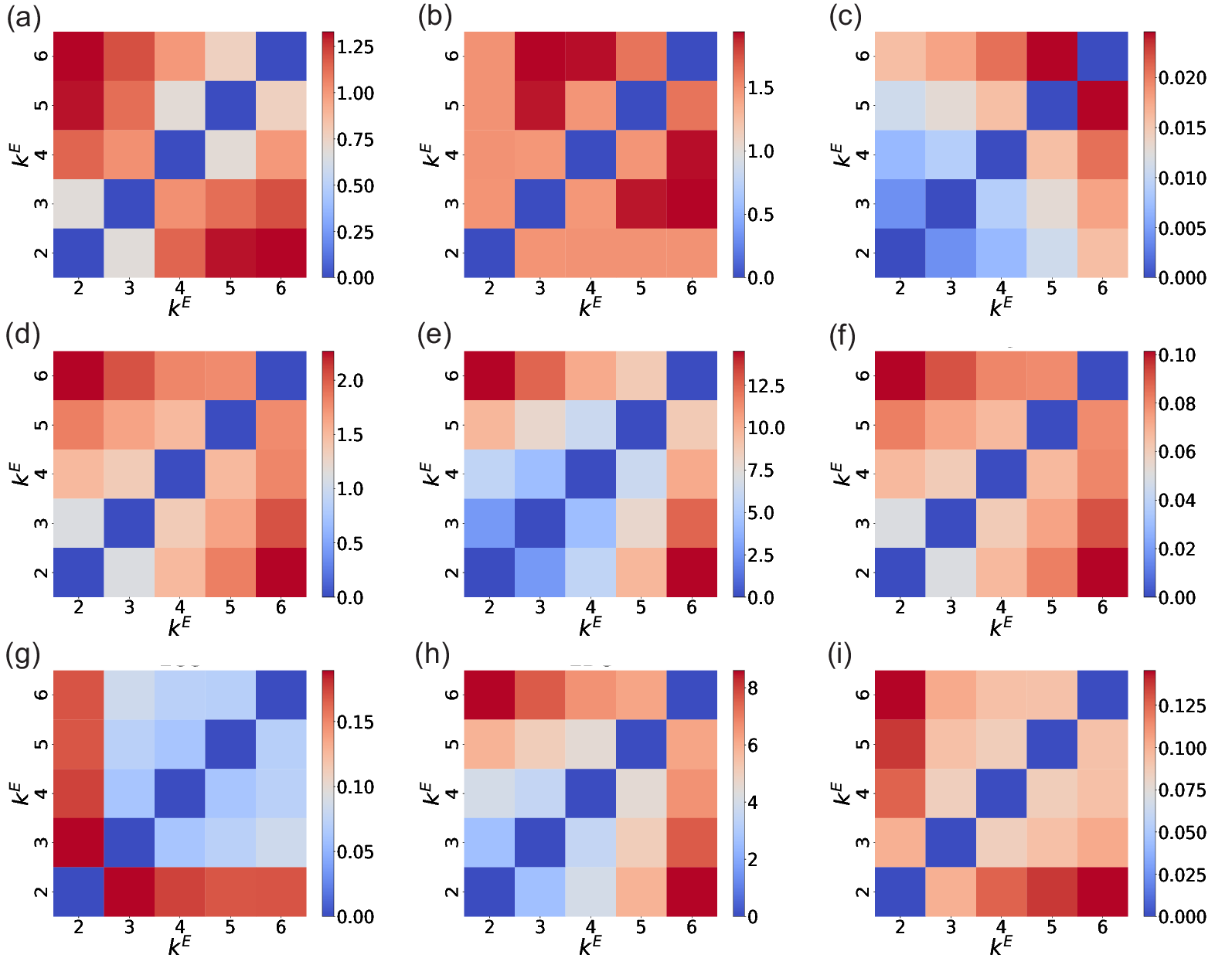}
    %\caption{节点数为1000，超边数为500的ERH网络差异性比较热力图}
    \caption{Comparison of hypernetworks generated by ERH, where $N=1000$ and $M=500$. The methods are: (a) HD; (b) Portrait; (c) Tensors; (d) HDC; (e) NDC; (f) VC; (g) ECC; (h) EDC; (i) HCC.}
    \label{fig:ERHheatmap}
\end{figure}

\begin{figure}[!htp]
    \centering
    \includegraphics[width=\columnwidth]{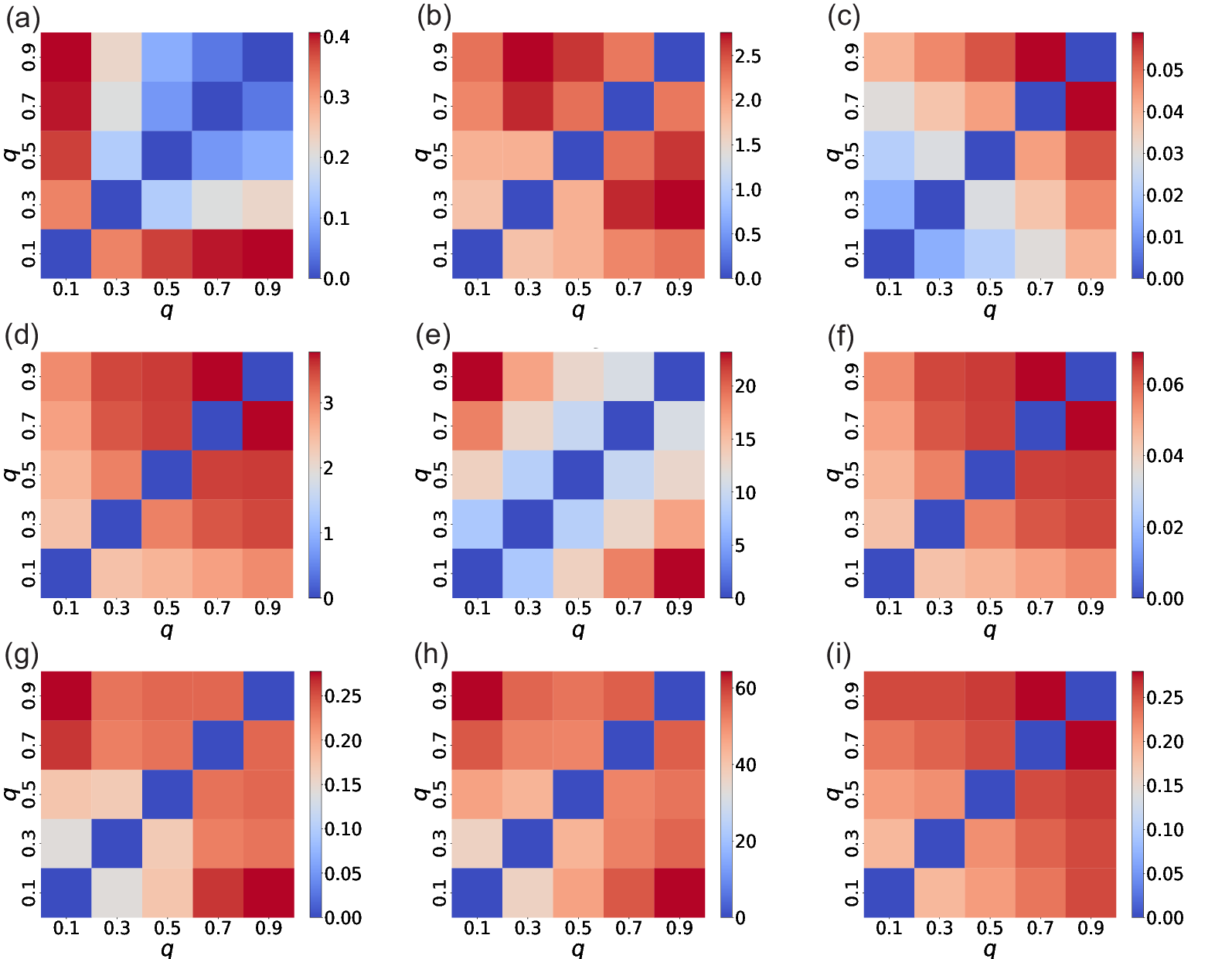}
    \caption{Comparison of hypernetworks generated by WSH, where $N=1000$ , $M=3000$ and $k^E=4$. The methods are: (a) HD; (b) Portrait; (c) Tensors; (d) HDC; (e) NDC; (f) VC; (g) ECC; (h) EDC; (i) HCC.}
    \label{fig:WSHheatmap}
\end{figure}

\begin{figure}[!htp]
    \centering
    \includegraphics[width=\columnwidth]{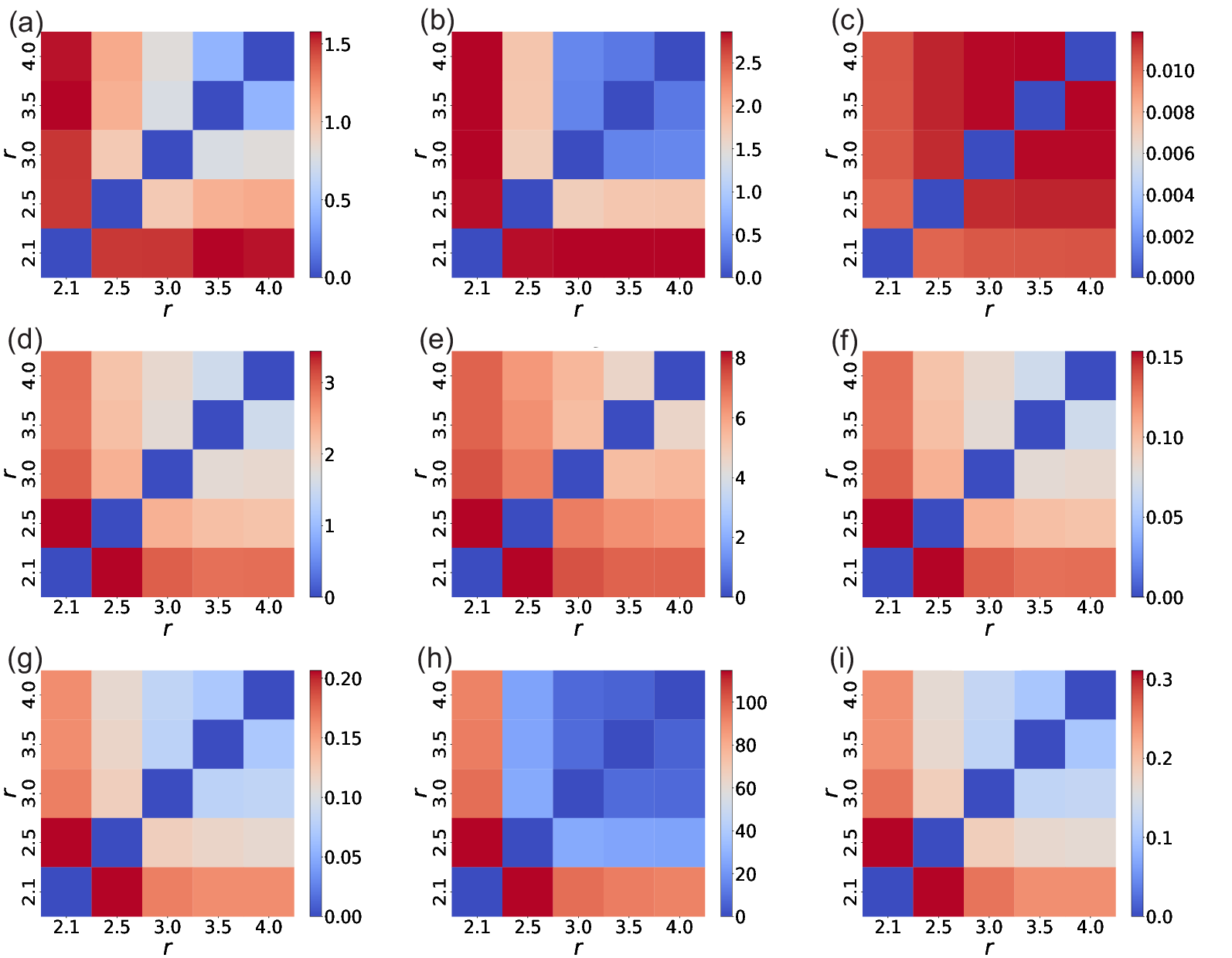}

    \caption{Comparison of hypernetworks generated by SFH, where $N=1000$ , $M=500$ and $k^E=4$. The methods are: (a) HD; (b) Portrait; (c) Tensors; (d) HDC; (e) NDC; (f) VC; (g) ECC; (h) EDC; (i) HCC.}
    \label{fig:SFHheatmap}
\end{figure}

The results of comparing uniform hypernetworks generated by ERH for various methods are depicted in Figure~\ref{fig:ERHheatmap}, with $k^{E}$ representing the size of hyperedges. For HD, we observe that hypernetworks are similar when the values of $k^{E}$ are close, which is in agreement with the way the hypernetworks are generated. The other baseline methods (except Portrait and ECC) show similar trends with those of HD with different values, indicating that the majority of the methods can distinguish hypernetworks generated by ERH with different $k^{E}$. We analyze the hypernetworks created by WSH and compare them in Figure~\ref{fig:WSHheatmap} based on different rewiring probabilities $q$. The results demonstrate that HD, Tensors, EDC, HCC, and NDC are capable of distinguishing hypernetworks created by different $q$ values. On the other hand, the other methods cannot. For instance, the dissimilarity between networks generated by $q=0.1$ and  $q=0.5$ and networks generated by $q=0.3$ and  $q=0.5$ are roughly the same, which is counter-intuitive. For the uniform hypernetworks generated by SFH, we show the difference between networks generated by different values of $r$ (see Figure~\ref{fig:SFHheatmap}), which is the slope of the degree distribution. It can be intuitively assumed that hypernetworks should be alike when the value of $r$ is near, and HD can demonstrate this pattern. In contrast, the other baselines are unable to distinguish between hypernetworks created with different $r$ values. Overall speaking, HD is robust to distinguish hypernetworks generated by different models, but the baselines have weakness in different scenarios.

\newpage
\clearpage

\begin{figure}[!htp]
    \centering
    \includegraphics[width=0.8\columnwidth]{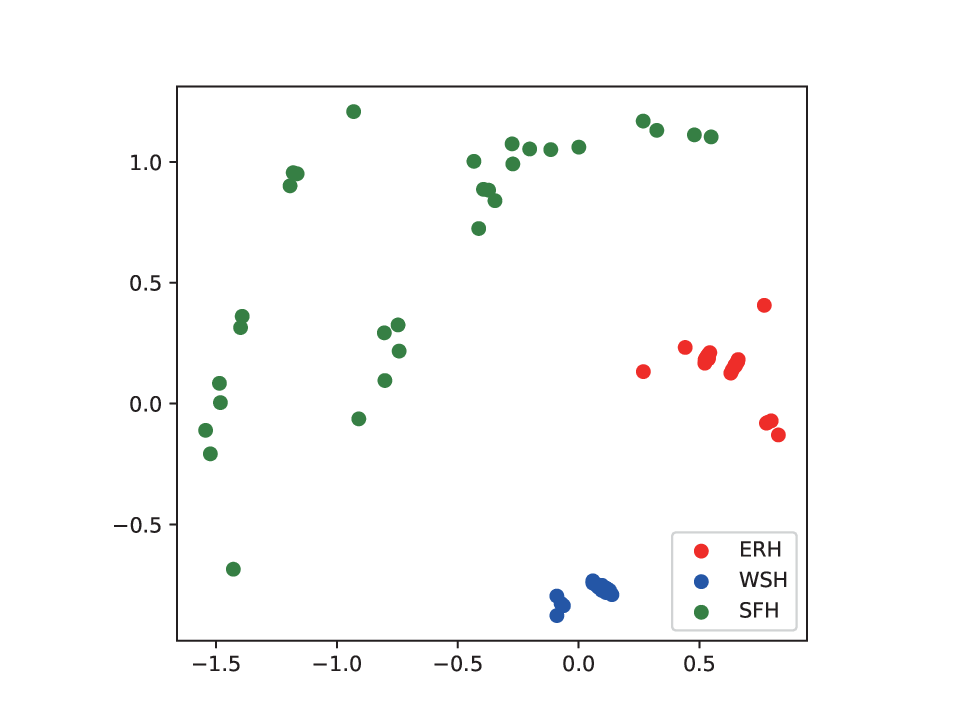}
    \caption{Synthetic hypernetwork classification: Red: ERH ($N=1000, M=500, k^E=4$), Blue: SFH ($N=1000, M=500, k^E=4, r=2.1$), Green: WSH ($N=1000, M=3000, k^E=4, q=0.05$).}
    \label{fig:MDS}
\end{figure}

We further explore whether our method can classify hypernetworks generated by different mechanisms and the results are given in Figure~\ref{fig:MDS}. We generate ERHs, WSHs, and SFHs with a network size of $N=1000$. For each type, we generate $30$ hypernetworks. The similarities between the hypernetworks are calculated by HD, and we further adopt multidimensional scaling map to show the distance between them in a geometric space. A shorter distance indicates that the two hypernetworks are more similar. The results of our analysis demonstrate that HD is effective in classifying hypernetworks, as those generated by the same mechanism are grouped together and those generated by different mechanisms are separated.

\subsection{Comparison of empirical hypernetworks}
We use six hypernetworks created from empirical data to prove the efficiency of our approach. Yuecai and Chuancai are recipe hypernetworks in which the hyperedges are made up of the components of each dish. Restaurants-Rev and Bar-Rev are review hypernetworks obtained from \url{Yelp.com}. These hypernetworks are composed of users who have reviewed the same restaurant or bar. NDC-classes is a hypernetwork of drugs, with each hyperedge representing a combination of class labels that make up a single drug. Algebra is collected from \url{MathOverflow.net}, where users who answered the same question are enclosed in a hyperedge. The information regarding the topological characteristics of the hypernetworks is presented in Table~\ref{table_data}. This table includes the number of nodes, the number of hyperedges, the degree, the average hyperdegree, the average size of the hyperedges, the clustering coefficient, the average shortest path length, the diameter, and the average link density of each hypernetwork. It is worth noting that the clustering coefficient, the average shortest path length, the diameter, and the average link density are computed based on the ordinary networks of the hypernetworks.

\begin{table}[!t]
\caption{The topological properties of hypernetworks, where $N$, $M$, $\langle k \rangle$, $\langle k^H \rangle$, $\langle k^E \rangle$, $c$, $\langle l \rangle$, $d$ and $\rho$ denote the number of nodes, the number of hyperedges, the average degree, the average hyperdegree, the average size of the hyperedges, the clustering coefficient, the average shortest path length, the diameter and the average density of hypernetwork, respectively.}
\begin{center} 
\label{table_data}
\begin{tabular}{cccccccccc} 
\toprule   
%   Data & $N$ & $M$ & $deg$ & $\langle k^H \rangle$ & $\langle k^E \rangle$ & $C$ & $\langle l \rangle$ & $d$ & $\rho$ \\  

 Data & $N$ & $M$ & $\langle k \rangle$ & $\langle k^H \rangle$ & $\langle k^E \rangle$ & $C$ & $\langle l \rangle$ & $d$ & $\rho$ \\ \midrule  
  Yuecai & 497 & 867 & 4.52 & 3.17 & 1.82 & 0.39 & 3.44 & 8 & 0.01   \\  
  Chuancai & 438 & 835 & 5.42 & 3.40 & 1.79 & 0.35 & 3.42 & 8 & 0.01   \\
  Restaurants-Rev & 565 & 601 & 8.14 & 8.14 & 7.66 & 0.54 & 1.98 & 5 & 0.14   \\
  Bars-Rev & 1234 & 1194 & 174.30 & 9.62 & 9.93 & 0.58 & 2.10 & 6 & 0.14\\
  NDC-classes & 1161 & 1088 & 10.72 & 5.54 & 5.92 & 0.61 & 3.50 & 9 & 0.01   \\
  Algebra & 423 & 1268 & 78.90 & 19.53 & 6.52 & 0.79 & 1.95 & 5 & 0.19\\
  \bottomrule 
\end{tabular}
\end{center}
\end{table}

\begin{figure}[!htp]
    \centering
    \includegraphics[width=\columnwidth]{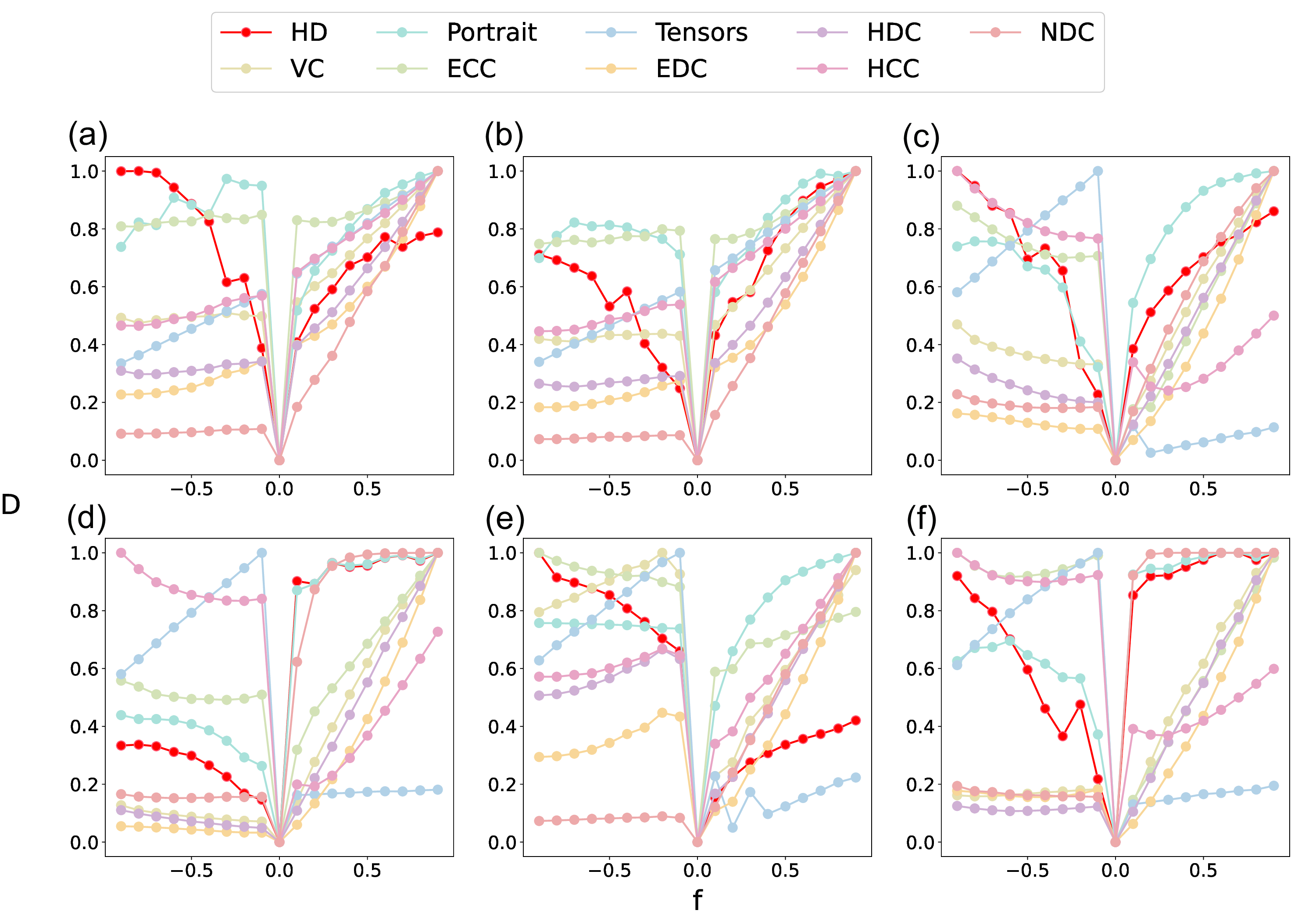}
    \caption{Results of comparing the differences in perturbation across different methods for different real networks: (a)Yuecai; (b)Chuancai; (c)Restaurants-Rev; (d)Bars-Rev; (e)NDC-classes; (f)Algebra.}
    \label{fig:perturbation}
\end{figure}

We assess the effectiveness of our proposed technique by performing hyperedge perturbations on empirical hypernetworks. We randomly add or take away a certain proportion ($f\in[-0.9,0.9]$) of hyperedges from a hypernetwork. A positive value of $f$ indicates the addition of hyperedges, while a negative value implies the deletion of hyperedges. The size of each added hyperedge is randomly chosen from the range of $[2, \xi]$, where $\xi$ is the maximum hyperedge size. We compare the difference between the original hypernetwork and the perturbed hypernetworks, the results are shown in Figure~\ref{fig:perturbation}. The proposed method, i.e., HD, exhibits a good increasing trend of the dissimilarity values in both hyperedge deletion and addition perturbations in all the empirical hypernetworks. It follows that if we make more changes to the number of hyperedges in the hypernetwork, the resulting hypernetworks will be more different from the original one, which is in line with our expectations. In contrast, baseline methods are inferior to HD. For example, the portrait method shows a clear decreasing trend of dissimilarity values in hyperedge deletion for hypernetworks such as Yuecai, Chuancai, and Algebra, which means if we delete more hyperedges from a hypernetwork, the generated hypernetwork is more similar to the original one.  Similarly, the Tensors method shows a decreasing trend in hyperedge deletion for all six hypernetworks, while the increasing trend is not significant in the hyperedge addition for Restaurant-Rev, Bars-Rev, NDC-classes, and Algebra. Moreover,  the six centrality methods, namely, ECC, EDC, HCC, HDC, NDC, and VC, show either an indistinct increasing or a decreasing trend in hyperedge deletion for the six hypernetworks.

\clearpage
\subsection{Parameter Analysis}

\begin{figure}[!htp]
    \centering
    \includegraphics[width=\columnwidth]{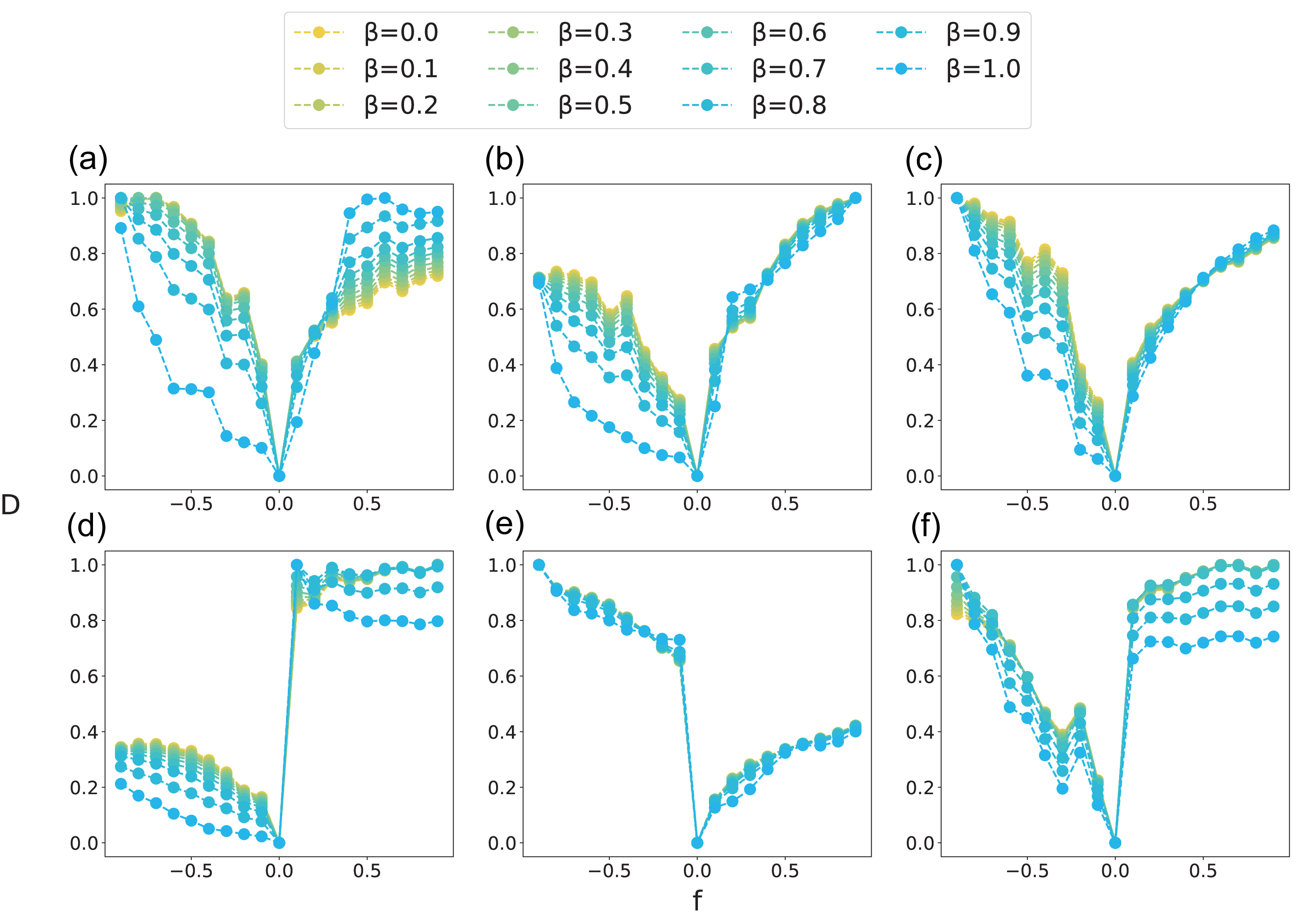}
    \caption{Comparing the differences in perturbations for different hypernetworks by altering the value of $\beta$: (a)Yuecai; (b)Chuancai; (c)Restaurants-Rev; (d)Bars-Rev; (e)NDC-classes; (f)Algebra.}
    \label{fig:weight}
\end{figure}

The method we proposed contains a parameter $\beta$ that emphasizes the importance of global and local dissimilarity between two hypernetworks. We further explore how the value of HD changes with $\beta$. The results are given in Figure~\ref{fig:weight} for six different hypernetworks. We observe that the values of HD with higher $\beta$ values are lower for hyperedge deletion process. However, for the hyperedge addition process, there is no consistent pattern of how HD changes with $\beta$. 

\clearpage
\section{Conclusions \& Discussion}
\label{Conclusions}
In this paper, we propose a method called the hyper-distance-based network comparison method (HD) to compare the dissimilarity between hypernetworks. 
HD is based on higher-order distances and utilizes node high-order distance distributions and Jensen-Shannon divergence. Specifically, we first compute the $s$-distance between each pair of nodes using s-walk and generate the $s$-distance distribution matrices of the hypernetwork. Next, we compare two hypernetworks via the Jensen-Shannon divergence, which encodes comparing the difference of the average connectivity between nodes at different $s$-distances and the heterogeneity of distance distributions. We compare the proposed HD method with other baselines on various synthetic and real hypernetworks. The experimental results demonstrate that HD is superior to the others in revealing the dissimilarity between hypernetworks. Future research directions include further expanding and improving the higher-order-distance-based method for comparing hypernetwork dissimilarity to adapt to more complex and large-scale network datasets. Furthermore, it can be extended to other types of networks such as multilayer hypernetworks and temporal hypernetworks~\cite{fischer2020visual,holme2012temporal}.

% \section*{Appendix}

% ...
% \section{CRediT authorship contribution statement}
% Tao Xu: Conceptualization, Methodology, Software, Data curation, Writing – original draft. Changgui Gu: Writing – review & editing, Funding acquisition. Xiu-Xiu Zhan: Supervision, Investigation.
% Ming Tang: Conceptualization, Methodology, Supervision.
\section{Declaration of competing interest}
The authors declare that they have no known competing financial interests or personal relationships that could have appeared to influence the work reported in this paper.
\section{Data availability}
Data will be available on request.
\section{Acknowledgments}
This work was supported by Natural Science Foundation of Zhejiang Province (Grant No. LQ22F030008), the National Natural Science Foundation of China (Grant No. 92146001), the Fundamental Research Funds for the Central Universities of Ministry of Education of China, and the Scientific Research Foundation for Scholars of HZNU (2021QDL030).

\bibliography{TempExample}

\end{document}